\documentclass{emulateapj}
\shorttitle{Fast orbital shrinkage of BHXBs driven by circumbinary disks}
\shortauthors{Chen \& Podsiadlowski}

\begin{document}


\title{Fast orbital shrinkage of black hole X-ray binaries driven by circumbinary disks }


\author{Wen-Cong Chen$^{1,2,3}$, and Philipp Podsiadlowski$^{2,3}$ }
\affil{$^1$ School of Physics and Electrical Information, Shangqiu Normal University, Shangqiu 476000, China;\\
 $^2$ Department of Physics, University of Oxford, Oxford OX1 3RH, UK;\\
 $^3$ Argelander-Insitut f\"{u}r Astronomie, Universit\"{a}t Bonn, Auf dem H\"{u}gel 71, 53121 Bonn, Germany;
chenwc@pku.edu.cn
}



\begin{abstract}
Recently, the black hole X-ray binary (BHXB) Nova Muscae 1991 has been reported to be experiencing an extremely rapid orbital decay. So far, three BHXBs have anomalously high orbital period derivatives, which can not be interpreted by the standard stellar evolution theory. In this work, we investigate whether the resonant interaction between the binary and a surrounding circumbinary (CB) disk could produce the observed orbital period derivatives. Analytical calculations indicate that the observed orbital period derivatives of XTE J1118+480 and A0620-00 can originate from the tidal torque between the binary and a CB disk with a mass of $10^{-9}~\rm M_{\odot}$, which is approximately in agreement with the dust disk mass detected in these two sources. However, Nova Muscae 1991 was probably surrounded by a heavy CB disk with a mass of $10^{-7}~\rm M_{\odot}$. Based on the CB disk model and the anomalous magnetic braking theory, we simulate the evolution of the three BHXBs with intermediate-mass donor stars by using the MESA code. Our simulated results are approximately consistent with the observed donor star masses, orbital periods, and orbital-period derivatives. However, the calculated effective temperatures of the donor stars are higher than indicated by the observed spectral types of two sources.
\end{abstract}


\section{Introduction}
Stellar mass black holes (BHs) are products of collapsing massive stars after they exhausted all nuclear fuel. Due to the ultra-strong gravitational field, anything including particles and electromagnetic radiation can not escape from inside of BHs.
Therefore, the best objects detecting BHs are X-ray binaries where the dynamical masses of BHs can be
estimated. At present, there exist two dozen BH candidates that have been identified in X-ray binaries \citep[][for a review]{remi06,casa14}. Most of them (19 sources) have been defined as BH low-mass X-ray binaries (BHLMXBs) because their donor star masses are less than 1 $\rm M_{\odot}$. Study of BHLMXBs will be of importance in understanding astrophysical process associated with ultra-strong gravitational fields, stellar and binary evolution, and common envelope (CE) evolution \citep[][for a review]{li15}.

In a standard CE model, it is difficult for low-mass donor stars to eject the massive envelope of BH progenitors during the CE phase \citep{port97,pods03}. As a result, the population synthesis predicted a birth rate to be two orders of magnitude lower than that derived from observations \citep{li15}. This difference can be solved by adopting an anomalously high CE efficiency parameter ($\alpha_{\rm CE}$) \citep{kalo99,yung08,kiel06}. As an alternative evolutionary channel, BHLMXBs may have evolved from BH intermediate-mass X-ray binaries driven by the anomalous magnetic braking of Ap/Bp stars \citep{just06} or surrounding circumbinary disks \citep{chen06}. Recently, \cite{wang16} found that BHLMXBs can be formed if most BHs are produced through a failed supernovae mechanism, in which the BH mass is equal to that of the He or CO core mass of the progenitor.

In the standard theory forming BHLMXBs, the angular-momentum-loss mechanisms usually include three cases as follows:
gravitational radiation, magnetic braking \citep{verb81}, and mass loss \citep{rapp82}. Therefore, orbital-period derivatives measured in some BHLMXBs can provide some valuable hints on their progenitors' evolution. Recently, the orbital-period derivatives of three BHLMXBs: XTE J1118 (hereafter 1118),  A0620-00 (hereafter 0620), and Nova Muscae 1991 (hereafter 1991) have been detected. \cite{gonz12} reported that 1118 is experiencing a rapid orbital shrinking at a rate $\dot{P}=-1.83\pm0.66~\rm ms\,yr^{-1}$. Subsequently, 0620 was also observed to have a negative orbital-period derivative of $\dot{P}=-0.6\pm0.1~\rm ms\,yr^{-1}$, and the orbital-period derivative of 1118 is refined to be $\dot{P}=-1.90\pm0.57~\rm ms\,yr^{-1}$ \citep{gonz14}.
In 2017, 1991 was detected be experiencing an extremely rapid orbital decay at a rate $\dot{P}=-20.7\pm12.7~\rm ms\,yr^{-1}$, which is significantly faster than those of 1118 and 0620 \citep{gonz16}.

\section{Analysis for the orbital evolution of BHLMXBs}
The orbital-angular momentum of a BHLMXB is $J=\Omega a^{2}M_{\rm bh}M_{\rm d}/(M_{\rm bh}+M_{\rm d})$, where $a$  is the orbital separation, $\Omega$ the orbital angular velocity of the binary, $M_{\rm bh}$, and $M_{\rm d}$ are the BH mass, and the donor star mass, respectively. Differentiating this equation, the change rate of the orbital period is
\begin{equation}
\frac{\dot{P}}{P}=3\frac{\dot{J}}{J}-3\frac{\dot{M}_{\rm d}}{M_{\rm d}}(1-q\beta)+\frac{\dot{M}_{\rm bh}+\dot{M}_{\rm d}}{M_{\rm bh}+M_{\rm d}},
\end{equation}
where $\beta=-\dot{M}_{\rm bh}/\dot{M}_{\rm d}$ is the BH accreting efficiency, $q=M_{\rm d}/M_{\rm bh}$ is the mass ratio of the binary. According to the first and the third term on the right hand side of Equation (1), the orbital-angular-momentum loss and the mass loss of the system can cause the orbit to shrink. However, the second term would produce a positive orbital-period derivative if material transferred from the less massive donor star to the more massive BH. In general, the angular-momentum-loss rate of BHLMXBs is $\dot{J}=\dot{J}_{\rm gr}+\dot{J}_{\rm mb}+\dot{J}_{\rm ml}+\dot{J}_{\rm ot}$, where $\dot{J}_{\rm gr}, \dot{J}_{\rm mb}, \dot{J}_{\rm ml}, \dot{J}_{\rm ot}$ represent the angular-momentum-loss rate caused by gravitational radiation, magnetic braking, mass loss, and other mechanisms, respectively.

Table 1 lists the relevant observed parameters of the three BHLMXBs. The orbital-period-change rate originating from gravitational radiation is
\begin{equation}
\dot{P}_{\rm gr}=-\frac{96G^{3}}{5c^{5}}\frac{M_{\rm bh}M_{\rm d}(M_{\rm bh}+M_{\rm d})}{a^{4}}P,
\end{equation}
where $G$ is the gravitational constant, $c$ the light velocity in vacuo. According to Equation (2), the orbital-period derivatives produced by gravitational radiation for 1118, 0620, 1991 are respectively $\sim 3.0, 2.0, 4.0\times 10^{-13}~\rm s\,s^{-1}$, which are obviously 2$-$3 orders of magnitude lower than the observed results.

Based on the standard magnetic braking prescription given by \cite{rapp83}, the corresponding orbital-period
derivative can be estimated to be
\begin{eqnarray}
\dot{P}_{\rm smb}=-1.4\times 10^{-12}\left(\frac{\rm M_{\odot}}{M_{\rm bh}}\right)\left(\frac{M_{\rm bh}+M_{\rm d}}{\rm  M_{\odot}}\right)^{1/3}\nonumber \\
\left(\frac{R_{\rm d}}{\rm R_{\odot}}\right)^{\gamma}\left(\frac{\rm d}{P}\right)^{7/3}~\rm s\,s^{-1},
\end{eqnarray}
where $R_{\rm d}$ is the donor-star radius. Adopting $\gamma=1$, the orbital-period derivatives given by magnetic braking are $\sim 7.8, 3.8, 2.2\times 10^{-12}~\rm s\,s^{-1}$ for 1118, 0620, 1991, respectively. These estimations are still one order of magnitude lower than these observed.
Actually, 1118 should have a fully convective donor star, which is not generally thought to produce magnetic braking \citep{rapp83,spru83}\begin{table*}
\centering
\begin{minipage}{170mm}
\caption{Some binary parameters of three BHLMXBs. The meaning of the columns are presented as follows: sources name, BH mass, donor star mass, donor star radius, orbital period, orbital separation, observed orbital period derivative, donor star spectrum type, and references.}
\begin{tabular}{lllllllll}
  \hline\hline\noalign{\smallskip}
Sources &$M_{\rm bh}$ & $M_{\rm d}$   & $R_{\rm d}$ & $P$ &  $a$ & $\dot{P}$& donor star &References\\
        & ($\rm M_{\odot})$ & ($\rm M_{\odot}$)& ($\rm R_{\odot}$) & (d) & ($\rm R_{\odot}$)  &($10^{-11}\rm s\,s^{-1}$)&spectrum type & \\
 \hline\noalign{\smallskip}
XTE J1118+480           &$7.46^{+0.34}_{-0.69}$ &$0.18\pm0.06$& $0.34\pm0.05$&0.1699  &$2.54\pm0.06$ &$-6.01\pm1.81$ & K5/M1 V& $1-6$\\
A0620-00                &$6.61^{+0.23}_{-0.17}$ &$0.40\pm0.01$& $0.67\pm0.02$  &0.3230&$3.79\pm0.04$ &$-1.90\pm0.26$  &K4 V& $6-10$  \\
Nova Muscae 1991        &$11.0^{+2.1}_{-1.4}$   &$0.89\pm0.18$& $1.06\pm0.07$&0.4326  &$5.49\pm0.32$  &$-65.6\pm40.3$  & K33/5 V & $11-16$ \\
\noalign{\smallskip}\hline
\end{tabular}
\end{minipage}
\\ \textbf{References}. (1) \cite{wagn01}; (2) \cite{mccl01}; (3) \cite{torr04}; (4) \cite{gonz08}; (5) \cite{calv09}; (6) \cite{gonz14}; (7) \cite{mccl86}; (8) \cite{oros94}; (9) \cite{gonz10}; (10) \cite{gonz11}; (11) \cite{remi92}; (12) \cite{casa97}; (13) \cite{oros96}; (14) \cite{wu15}; (15) \cite{wu16};  (16) \cite{gonz16}.
\end{table*}

To account for the formation of compact BHLMXBs, \cite{just06} proposed an anomalous magnetic braking (AMB) mechanism, which is caused by the coupling between the strong magnetic field of Ap/Bp stars and an irradiation-driven wind induced by the X-ray flux.
The orbital-period derivative predicted by the AMB model is given by
\begin{eqnarray}
\dot{P}_{\rm amb}=-2.4\times 10^{-8}\left(\frac{B_{\rm s}}{10000\rm G}\right)\left(\frac{M_{\rm bh}+M_{\rm d}}{M_{\rm bh}}\right)\nonumber \\
\left(\frac{f}{0.001}\frac{\dot{M}_{\rm bh}}{10^{-9}M_{\odot}\rm yr^{-1}}\right)^{0.5}\left(\frac{R_{\rm d}}{R_{\odot}}\right)^{15/4}\nonumber \\
\left(\frac{R_{\odot}}{a}\right)^{2}\left(\frac{M_{\odot}}{M_{\rm d}}\right)^{7/4}\left(\frac{P}{\rm d}\right)~\rm s\,s^{-1},
\end{eqnarray}
where $B_{\rm s}$ is the surface magnetic field of the donor star, $f$ is the wind-driving efficiency. According to the equation given by \cite{king96}, the accretion rate of BHs can be estimated to be $\dot{M}_{\rm bh}\sim 0.1, 0.5, 2.0\times 10^{-9}~M_{\odot}\rm yr^{-1}$ for 1118, 0620, and 1991, respectively \citep{gonz16}. Assuming a wind-driving efficiency of $f=0.001$ and a surface magnetic field of $B_{\rm s}=5000$ G, the resulting orbital-period derivatives are $\dot{P}\sim 5.3, 6.1, 3.5\times 10^{-11}~\rm ss^{-1}$ for 1118, 0620, and 1991, respectively. Even if taking such an ultra-strong field of 5000 G, the orbital-period derivative induced by AMB mechanism it still one order of magnitude lower than that of 1991.

Therefore, it seems that there are other efficient angular-momentum-loss mechanisms to cause the rapid orbital decay of the three BHLMXBs. Dramatically, \cite{muno06} have detected that the excess mid-infrared-emission area
are obviously larger than the binary-orbit areas these systems, and they suggested that it probably arise from a contribution of circumbinary (CB) disks. Recently, observations performed by the Wide-Field Infrared Survey Explorer have confirmed that these two sources should be surrounded by CB disks \citep{wang14}. In this work, we attempt to explore whether a CB disk around these three sources could be responsible for their observed orbital period derivatives. In section 3, we describe the CB disk model, and constrain the CB disk masses. In Section 4, we use the MESA code to simulate the formation of the three BHLMXBs. Finally, we summarize the results with a brief conclusion and discussion in Section 5.

\section{CB disk model}

In this section, we investigate whether the rapid orbital-decay observed in the three BHLMXBs could be interpreted
by CB disks around these sources. The resonant theory between a binary and its CB disk is based on a standard thin disk \citep{gold79,arty94}, in which $H/R=0.01-0.1$ ($H$, and $R$ are the thickness and the half angular momentum radius of the CB disk, respectively).
This resonant torque can be estimated using the viscous torque of the CB disk, which can be written as the following relation \citep{lubo96,derm13}
\begin{equation}
\dot{J}_{\rm d}=M_{\rm cb}\Omega\nu,
\end{equation}
where $M_{\rm cb}$ is the CB-disk mass, $\nu=R^{2}(H/R)^{2}\alpha\Omega_{\rm d}$ is the disk viscosity ($\alpha$, and $\Omega_{\rm d}$ are the viscous parameter and the angular velocity of the CB disk, respectively).
Therefore, the orbital separation derivative of BHLMXBs is given by \citep{lubo96,derm13}
\begin{equation}
\frac{\dot{a}}{a}=-\frac{2l}{m}\frac{M_{\rm cb}\alpha}{\mu}\left(\frac{H}{R}\right)^{2}\frac{a}{R}\Omega,
\end{equation}
where $l$, and $m$ are the time-harmonic number and the azimuthal
number (Artymowicz \& Lubow 1994), and $\mu$ is the reduced mass of the binary.

Differentiating the Keplerian third law $G(M_{\rm bh}+M_{\rm d})/a^{3}=4\pi^{2}/P^{2}$, we can obtain the orbital-period derivative of BHLMXBs as follows,
\begin{equation}
\frac{\dot{P}}{P}=\frac{3\dot{a}}{2a}+\frac{\dot{M}_{\rm bh}+\dot{M}_{\rm d}}{2(M_{\rm bh}+M_{\rm d})}.
\end{equation}
Assuming that the mass-loss rate of BHLMXBs during the mass transfer is $\dot{M}_{\rm bh}+\dot{M}_{\rm d}=-1.0\times10^{-7}~\rm M_{\odot}\,yr^{-1}$ (an ultra-high mass-loss rate), and $M_{\rm bh}+M_{\rm d}=10~\rm M_{\odot}$, we can estimate the second term on the right hand side of
Equation (7) to be $-5\times 10^{-9}~\rm yr^{-1}$. For a binary with an orbital period of 0.5 d, the contribution of this term is
$\dot{P}\sim-0.2~\rm ms\,yr^{-1}$, which is obviously lower than those of the three BHLMXBs. Therefore, in this section we ignore the effect of the mass loss on the orbital-period derivative. Combining equations (6) and (7), and considering the resonances are very weak ($m=l$) when the eccentricity $e\leq 0.1\sqrt{\alpha}$ \citep{derm13}, the orbital-period derivative predicted by the CB disk model is
\begin{equation}
\dot{P}=-6\pi\frac{M_{\rm cb}\alpha}{R}\left(\frac{H}{R}\right)^{2}\frac{a}{\mu}.
\end{equation}

The inner radius of the CB disk should locate a distance to the mass center of the BHLMXBs as $r_{\rm in}=1.7a$, at which the disk would be tidally truncated \citep{taam03,dubu04}. In addition, the lack of excess flux at $24\mu\rm m$ in the observation for 1118 and 0620 imply that the outer radius of the disk is near $r_{\rm out}=3a$ \citep{muno06}. Therefore, we can obtain a half angular momentum radius to be $R=(r_{\rm in}+r_{\rm out})/4+\sqrt{r_{\rm in}r_{\rm out}}/2=2.3a$. Assuming that the CB disks in these three sources have the same relation between $R$ and $a$, Equation (8) reveals that the orbital-period derivative is related to two factors: a degenerate CB disk parameter ($\frac{M_{\rm cb}\alpha H^{2}}{R^{2}}$) and a binary parameter ($1/\mu$). Based on the observed masses of two components, and taking $H/R=0.1$, $\alpha=0.1$, we can constrain the CB disk mass for the three BHLMXBs.

 In Figure 1,  we compare the orbital-period derivatives predicted by the CB disk scenario with observations in the $\dot{P}-1/\mu$ diagram. \cite{muno06} provided an estimation ($\sim10^{-9}~\rm M_{\odot}$) for CB disk masses  surrounding 1118 and 0620. Since the CB disk masses of different BHLMXBs should have a dispersion, a two orders of magnitude mass range is considered. In Figure 1, the solid, dashed, and dotted curves represent the predicted $\dot{P}$ derived by equation (8) under a CB disk mass of $10^{-7}$, $10^{-8}$, and $10^{-9}~\rm M_{\odot}$, respectively. It is clear that the observed parameters of 1118 and 0620 are well fitted by the theoretical line of  $10^{-9}~\rm M_{\odot}$, which is the estimated CB disk masses of these two sources. For 1991, a relatively heavy CB disk ($\sim10^{-7}~\rm M_{\odot}$) would be expected in order to account for the observed orbital-period derivative.
\begin{figure}
\centering
\includegraphics[width=1.15\linewidth,trim={0 0 0 0},clip]{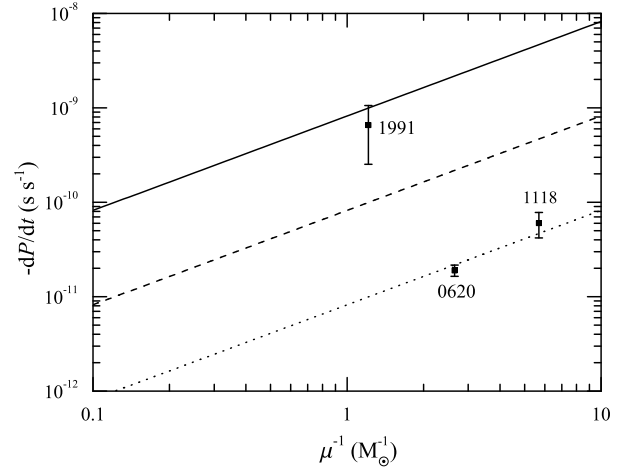}
\caption{Comparison of the predicted orbital-period derivatives by the CB disk scenario with observations in the $\dot{P}-1/\mu$ diagram. The solid squares denote the three BHLMXBs. The solid, dashed, and dotted curves represent the CB disk mass of $10^{-7}$, $10^{-8}$, and $10^{-9}~\rm M_{\odot}$ (the observed CB disk mass in 1118 and 0620), respectively. } \label{fig:orbmass}
\end{figure}

\section{Simulation of BHLMXBs}
\subsection{Input physics}
In this section, we use a \texttt{MESAbinary} update version (8118)
in \texttt{MESA} module \citep{paxt15} to simulate the formation of the three BHLMXBs.  The evolutionary beginning is assumed to be a binary system containing an intermediate-mass donor star (with a mass of $M_{\rm d}$) and a BH (with a mass of $M_{\rm bh}$). For the donor-star compositions, we adopt a solar compositions ($X = 0.70, Y = 0.28$, and $Z = 0.02$). Meanwhile, the two components are thought to be circularized at all times.

Once the donor star overflows its Roche lobe by a long-term nuclear evolution, the material would be transferred from the donor star to the BH through the inner Lagrangian point at a rate of $\dot{M}_{\rm tr}$. The accretion rate of the BH is limited to the Eddington rate as follows
\begin{equation}
\dot{M}_{\rm Edd}=2.6\times10^{-7}\frac{M_{\rm bh}}{10 M_{\odot}}\left(\frac{0.1}{\eta}\right)\left(\frac{1.7}{1+X}\right) M_{\odot}\,\rm yr^{-1},
\end{equation}
where $X$ is the hydrogen abundance in the accreting material, and
\begin{equation}
\eta=1-\sqrt{1-\left(\frac{M_{\rm bh}}{3M_{\rm bh,0}}\right)^{2}}
\end{equation}
is the energy conversion efficiency of the BH, where $M_{\rm bh,0}$ is the initial BH mass \citep[see also][]{bard70,king99}.
Therefore, the accretion rate of the BH is $\dot{M}_{\rm bh}={\rm min}[\dot{M}_{\rm Edd}, -\dot{M}_{\rm tr}]$. If the accretion process is super-Eddington, we assume that a constant fraction $\delta$ of the lost
mass feeds into the CB disk surrounding the BHXB, i. e. the mass increasing rate of the CB disk is
\begin{equation}
\dot{M}_{\rm cb}=-\delta(\dot{M}_{\rm tr}+\dot{M}_{\rm Edd}).
\end{equation}

Similar to \cite{chen16}, we consider the wind loss from the donor star is driven by X-ray irradiation. The irradiation-driving wind loss rate is given by
\begin{equation}
\dot{M}_{\rm w}=-f_{\rm ir}L_{\rm X}\frac{R_{\rm d}^{3}}{4 G M_{\rm d} a^{2}},
\end{equation}
where $f_{\rm ir}$ is the irradiation efficiency (in this work, we take $f_{\rm ir}=10^{-3}$). We calculate the X-ray luminosity by $L_{\rm X}=\eta\dot{M}_{\rm bh}c^{2}$. Therefore, the mass loss rate of the donor star is $\dot{M}_{\rm d}=\dot{M}_{\rm tr}+\dot{M}_{\rm w}$.

Assuming that 1118 originated from the Galactic disk and the donor has solar metallicity, \cite{frag09} found that this system includes a $\sim6.0-10.0~M_{\odot}$ BH and a $\sim1.0-1.6~M_{\odot}$ donor star. However, some clues indicate that an intermediate-mass ($\ga2.0~M_{\odot}$) should be a plausible range for the progenitor mass of the donor stars in BHLMXBs. First, it still remains controversial whether a donor star with a mass less than $\la1.5~M_{\odot}$ can provide sufficient orbital energy to eject the envelope of the black-hole progenitor \citep{pods95,port97,kalo99,pods03}. Second, CNO-processed elements were observed on the surface of 1118 \citep{hasw02}, which implies its progenitor should be an intermediate-mass star.

In the input physics calculating the evolution of binary stars, orbital angular-momentum losses are key issue. In the \texttt{MESA} code, we consider four types of orbital angular-momentum
loss during the evolution of BHXBs: (1) gravitational-wave radiation; (2)
anomalous magnetic braking: we adopt the same magnetic braking prescription
given by \cite{just06} and \cite{chen16}; (3) mass loss: the
mass loss from the vicinity of the BH is
assumed to be ejected in the form of isotropic winds and to carry away the specific orbital angular of the
BH, while the donor star winds carry away that of the donor star; (4) tidal torque produced by the interaction between the CB disk and the BHXB. According to Equation (5), the angular momentum loss rate extracting by the CB disk can be written as
\begin{equation}
\dot{J}_{\rm cb}=-M_{\rm cb}\alpha\left(\frac{H}{R}\right)^{2}\frac{a^{3}}{R}\Omega^{2}.
\end{equation}

\begin{figure}
\centering
\includegraphics[width=1.15\linewidth,trim={0 0 0 0},clip]{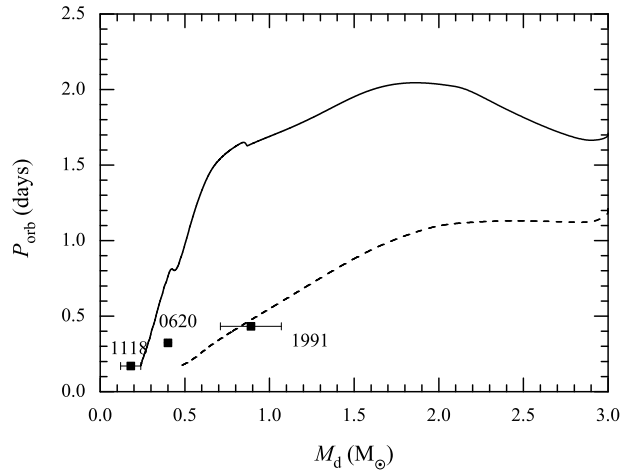}
\caption{Evolutionary tracks of BHXBs consisting of a donor star with a mass of $3.0~\rm M_{\odot}$ and a BH with a mass of $6.0~\rm M_{\odot}$ (solid curve) or $10.0~\rm M_{\odot}$ (dashed curve) in the $P_{\rm orb} - M_{ \rm d}$ diagram. The solid, and dashed curves denote an initial orbital period of 1.21, and 1.71 days, respectively. The solid squares represent the three observed BHLMXBs.} \label{fig:orbmass}
\end{figure}

\begin{figure}
\centering
\includegraphics[width=1.15\linewidth,trim={0 0 0 0},clip]{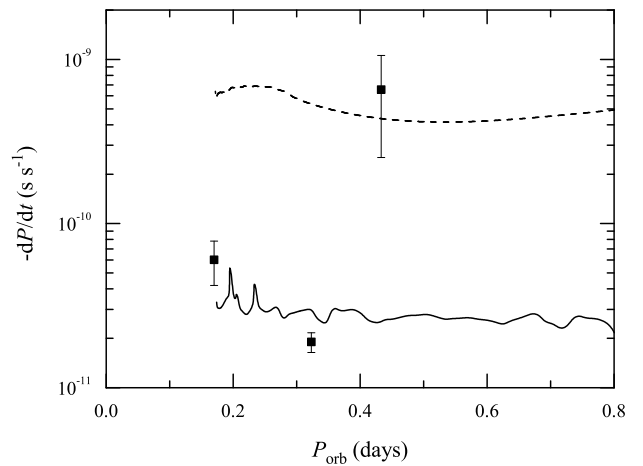}
\caption{Same as Figure 2, but for the $|\dot{P}| - P_{ \rm orb}$ diagram. The solid squares represent three observed BHLMXBs.} \label{fig:orbmass}
\end{figure}

\subsection{Results}
In our calculation, the donor stars in BHXBs are assumed to be Ap/Bp star with an initial mass of $3.0~\rm M_{\odot}$ and a surface magnetic field of 500 G, and the initial masses of BHs are $6.0~\rm M_{\odot}$ (for 1118 and 0620) and $10.0~\rm M_{\odot}$ (for 1991). To fit the CB disk mass inferred in Section 3, a faction $\delta=5.0\times10^{-9}$ and $5.0\times 10^{-7}$ of the mass loss during the super-Eddington accretion is thought to feed into the CB disk. By changing the initial orbital periods, we can diagnose whether the relevant BHXBs can evolve into the three observed sources by comparing the donor star masses, orbital periods, and orbital-period derivatives.

Our calculation show that the CB disk masses are approximately consistent with the inferred mass in Section 3 when the initial orbital-periods are 1.21 d and 1.71 d for 1118 and 1991, respectively. In Figure 2, we plot the evolution of BHXBs in the $P_{\rm orb} - M_{ \rm d}$ diagram. It is clear that 1118 and 0620 can evolved from a BHXB with an initial orbital-period $P_{\rm i}=1.21$ days, while the progenitor of 1991 should have an initial orbital-period $P_{\rm i}=1.71$ d. Figure 3 presents the evolution of orbital-period derivatives with the orbital periods. Both cases are approximately in agreement with the observed values of 1118 or 0620, and 1991.

\begin{figure*}
\centering
\begin{tabular}{cc}
\includegraphics[width=0.48\textwidth,trim={30 10 30 30},clip]{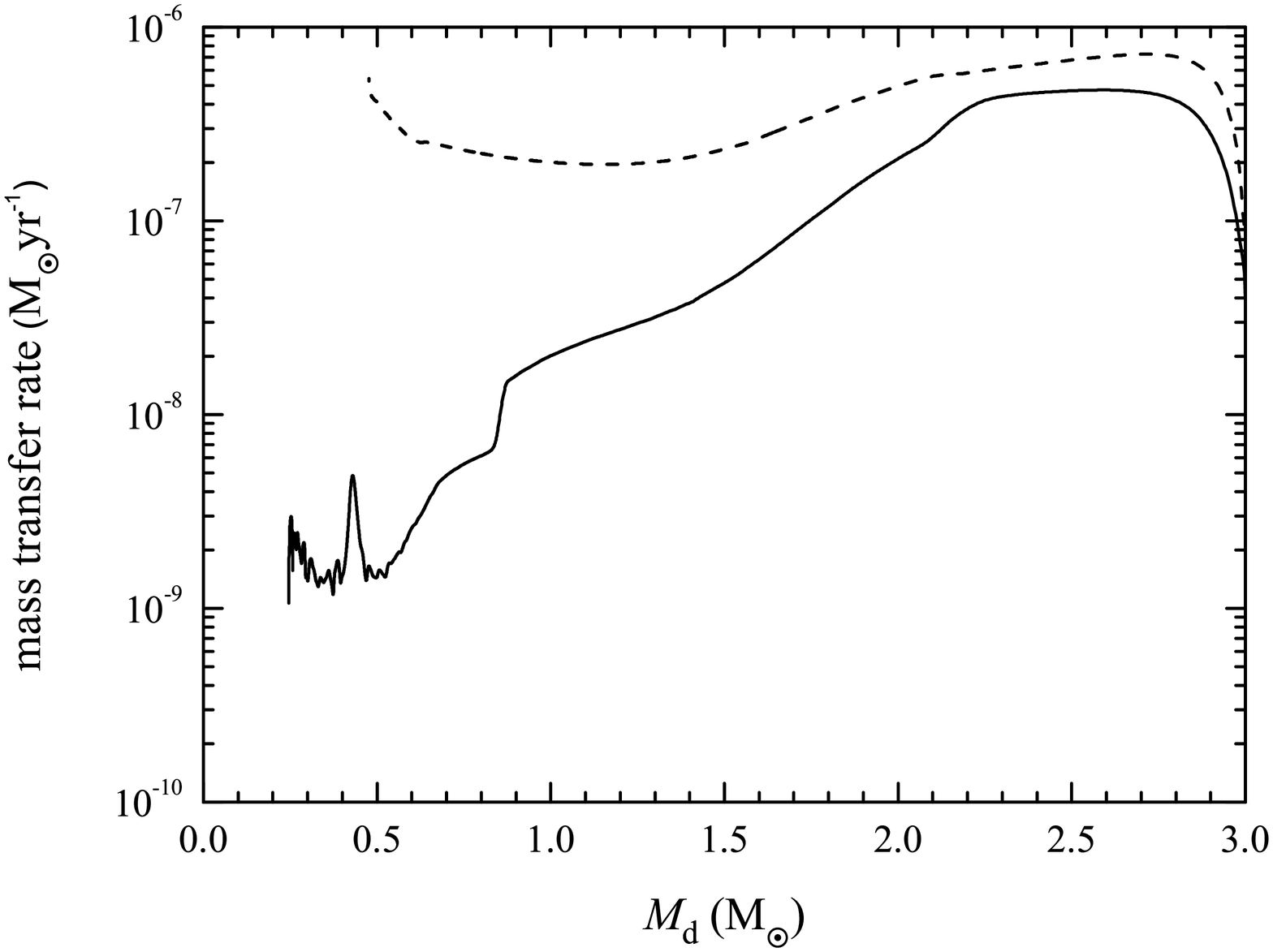} &    \includegraphics[width=0.48\textwidth,trim={30 10 30 30},clip]{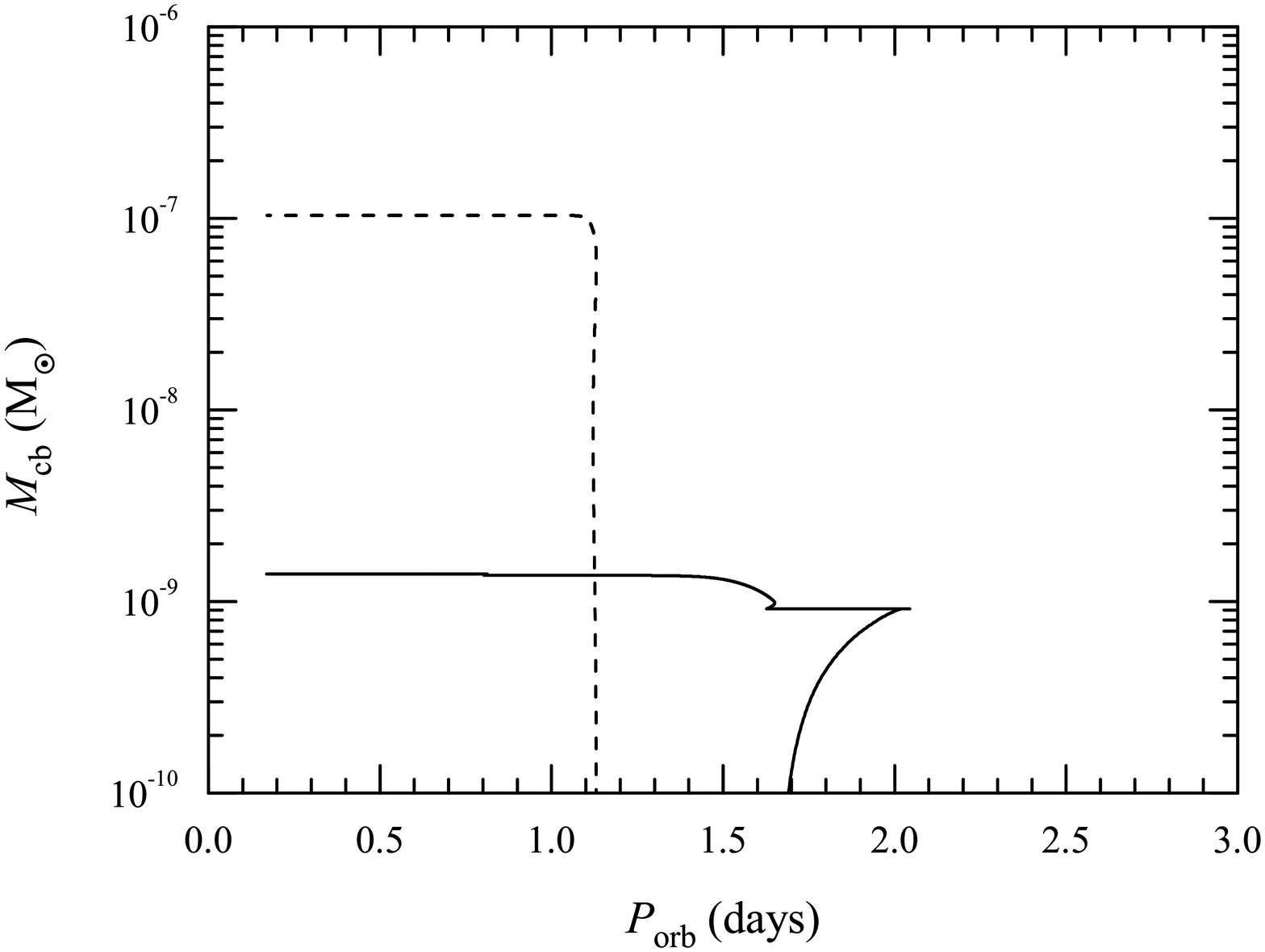} \\
\end{tabular}
\caption{\label{fig:pulses}Same as Figure 2, but for the evolution of the mass transfer rate in the $\dot{M}_{\rm d}-M_{\rm d}$ (left panel), and the evolution of the CB disk mass in the $M_{\rm cb}-P_{\rm orb}$ (right panel). }
\end{figure*}

Figure 4 summarizes the evolution of the mass transfer rate and the CB disk mass. When $P_{\rm i}=1.21$ days, the CB disk mass reaches a maximum of $\sim1.4\times10^{-9}~\rm M_{\odot}$ at the donor-star mass of $\sim0.4~\rm M_{\odot}$ (at this moment $P_{\rm orb}\approx 0.8$ d) due to the mass transfer of super-Eddington. Because the CB-disk mass increases very slowly, the orbital-period first increases, and then sharply deceases due to a relatively high disk mass (see also equation 13, results in an efficient angular-momentum loss) when $P_{\rm orb}\approx 2.0$ d. Subsequently, two short reversals of the orbital period also correspond to the two new CB-disk masses. For the case $P_{\rm i}=1.71$ days, the CB disk mass rapidly increases to a maximum of $M_{\rm cb}\sim1.0\times10^{-7}~\rm M_{\odot}$ when the orbital period is $\sim 1.1$ d and the donor-star mass is $1.9-2.0~\rm M_{\odot}$. At the current stage, 1118 and 1991 have an mass transfer rate of $\sim2\times10^{-9}$ and $\sim2\times10^{-7}~\rm M_{\odot}\, yr^{-1}$, respectively.

In figure 5, we also compare the simulated results with the effective temperatures indicated by the observed donor-star spectral types and orbital periods in the $T_{\rm eff}-P_{\rm orb}$ diagram. The effective temperature of the donor star in 0620 is approximately consistent with the simulated result. However, the donor stars in 1118 and 1991 were detected as cool spectral types. A similar problem had already been noticed by the previous works performed by \cite{just06} and \cite{chen06}.

\begin{figure}
\centering
\includegraphics[width=1.15\linewidth,trim={0 0 0 0},clip]{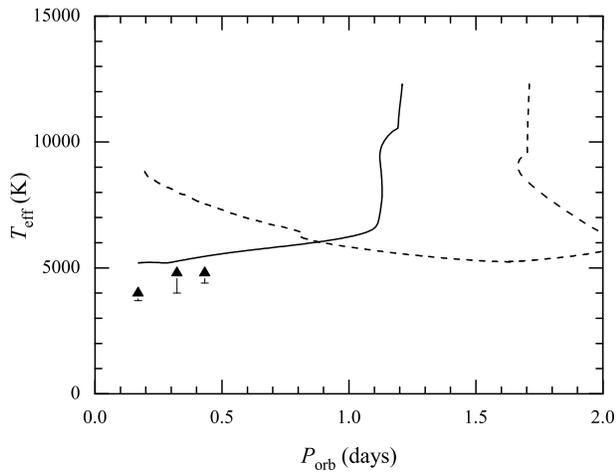}
\caption{Same as Figure 2, but for the evolution of the effective temperature of the donor star in the $T_{\rm eff} - P_{\rm orb}$ diagram.  } \label{fig:orbmass}
\end{figure}

\section{Discussion and Summary}
Recently, the three BHLMXBs including 1118, 0620, and 1991 were reported to be experiencing an extremely fast orbital decay.
The detected orbital-period derivatives are $1-3$ orders of magnitude higher than those given by gravitational radiation, and standard magnetic braking. For the AMB, the estimated $\dot{P}$ for 1991 is still one order of magnitude lower than observations even if the donor star has an ultra-strong magnetic field of 5000 G.

In this work, we attempt to explore whether the observed orbital decay can be interpreted by the existence of CB disks around BHLMXBs. Adopting some typical CB disk parameters $H/R=0.1$, and $\alpha=0.1$, the observed $\dot{P}$ in the three sources could be explained by a surrounding CB disk with a mass of $10^{-9}~\rm M_{\odot}$ (for 1118 and 0620) or $10^{-7}~\rm M_{\odot}$ (for 1991). Dramatically, the inferred CB disk masses are approximately consistent with the observed results in mid-infrared emission for 1118 and 0620 \citep{muno06}. The CB disks surrounding BHLMXBs may originate from three following channels: (1) CB disks are the remnants of the common envelope. In principle, compact binary systems should experience a common envelope evolutionary phase \citep{ivan13}. If the common envelope can not be fully ejected, the remaining material may collapse into a CB disk surrounding the binary system \citep{spru01}. (2) CB disks are the products of the mass transfer. A fraction of the mass loss probably form a disk structure surrounding the binary rather than leave it \citep{heuv74,heuv94}. (3) CB disks could be fed by mass loss during single outburst or successive outburst in BHLMXBs \citep{xu18}.

In this work, we also simulate the formation of the three BHLMXBs 1118, 0620,  and 1991 by using the MESA code. In the calculation, we assume that a fraction $\delta$ of the mass loss during super-Eddington accretion of BHXBs forms a CB disk surrounding the binary. To fit the CB disk mass inferred in Section 3, $\delta$ should be $5\times10^{-9}$ and $5\times10^{-7}$ for 1118 (or 0620), and 1991, respectively. Our simulations indicate that, the progenitor of 1118 (or 0620) may be a BH intermediate-mass X-ray binary consisting a 6.0 $M_{\odot}$ BH and a 3.0 $M_{\odot}$ donor star, and with an initial orbital period of 1.21 d; while the progenitor of 1991 should have a heavy BH (10.0 $M_{\odot}$), and a relatively wide orbit (initial orbital period is 1.71 d). Our simulated donor-star masses, the donor-star radii, the orbital periods, and the orbital-period derivatives are approximately in agreement with the observed results. For 1118, the calculated mass-transfer rate is $2\times10^{-9}~\rm M_{\odot}\, yr^{-1}$ at the current stage. However, the observed peak luminosity of 1118 is $\sim 10^{-3}~L_{\rm Edd}$ ($L_{\rm Edd}$ is the Eddington luminosity) \citep{wu10}, which implies the accretion rate of the BH is $\sim10^{-10}~\rm M_{\odot}\, yr^{-1}$. \cite{nara95} proposed that the critical rate $\dot{M}_{\rm crit}\sim\alpha^{2}\dot{M}_{\rm Edd}$ ($\alpha$ is the viscous parameter of the accretion disk)  for the advection-dominated accretion flow. If we take $\alpha=0.1$, then $\dot{M}_{\rm crit}\sim10^{-9}~\rm M_{\odot}\, yr^{-1}$, which is the same order of magnitude with our simulated mass-transfer rate. Therefore, the advected energy are probably lost into the BH, and the radiation efficiency of the accretion disk in 1118 is relatively low.

Recently, \cite{xu18} also employed the CB disk model to account for the fast orbital decay in these three sources. However, their initial donor star masses are 1.0 $M_{\odot}$, which is difficult to result in  CNO-processed elements detected on the surface of 1118 \citep{hasw02}. In addition, they assumed that the CB disk is formed due to single outburst or successive outburst at current time. Therefore, their model only produced a high orbital period derivative in a relatively short timescale.

Certainly, our simulation present a relatively high effective temperature of the donor stars. The main reasons could be as follows. First, \cite{torr04} found that the donor star of 1118 was only detected $\sim 55$ percent light during quiescence, hence the determination for the spectral types of the donor stars in such systems are controversial. Second, the irradiation process of X-ray could alter the effective surface boundary condition of the donor stars, especially change the ionization degree of the hydrogen at the bottom of the irradiate layer \citep{pods91}.

\acknowledgments {We thank the referee for his/her very careful reading and
comments that have led to the improvement of the
manuscript. This work was partly supported
by the National Natural Science Foundation of China (under grant number 11573016, and 11733009),
the Program for Innovative Research Team (in Science and Technology) at the
University of Henan Province, and the China Scholarship Council. This work has also been supported
by a Humboldt Research Award to PhP at the University
of Bonn.}

\end{document}